\documentclass{svproc}
\usepackage{url}
\usepackage{graphicx}
\usepackage{multicol}
\usepackage{footmisc}
\usepackage{caption}
\usepackage{amsmath}
\usepackage{bm}  %
\usepackage{ulem}
\usepackage{mwe}
\usepackage{comment}
\usepackage{verbatim}
\usepackage[titletoc,title]{appendix}
\usepackage{comment}
\usepackage{wrapfig}
\usepackage{lipsum}
\usepackage{floatrow}

\begin{document}
\mainmatter              
\title{Loss of conformality in Efimov Physics}
%
%
\author{Abhishek Mohapatra\inst{1,2} \and Eric Braaten\inst{2}
}
%
%
\tocauthor{Abhishek Mohapatra, and Eric Braaten}
\institute{Department of Physics, Duke University, Durham, NC 27705, USA,\\
\email{am749@duke.edu},
\and
Department of Physics, The Ohio State University, Columbus, OH 43201, USA,\\
\email{braaten.1@osu.edu}.}

\maketitle              

\begin{abstract}
The loss of conformal invariance in Efimov physics is due to the merger and disappearance of an infrared and an ultraviolet fixed point of a three-body renormalization group flow as the spatial dimension $d$ is varied. In the case of identical bosons at unitarity,  it is known that there are two critical dimensions $d_{\rm 1}=2.30$ and $d_{\rm 2}=3.76$  at which there is loss of conformality.  For $d<d_{\rm 1}$ and $d>d_{\rm 2}$, the beta function of the  three-body coupling  has real roots which correspond to infrared and ultraviolet fixed points.  The fixed points merge and disappear into the complex plane at the critical dimensions $d_1$ and $d_2$.
For $d_{\rm 1}<d<d_{\rm 2}$, the beta function has complex roots and the renormalization group flow for the three-body coupling constant is a limit cycle.
\end{abstract}
\section{Introduction}
In relativistic systems, the conformal symmetry group includes Poincar\'{e} symmetry and continuous scaling symmetry as subgroups. From the renormalization group (RG) perspective, scale invariance in any system arises at the fixed points of a RG flow. A general mechanism for the loss of conformal invariance in the system is the merging of an infrared (IR) and an ultraviolet (UV) fixed point and their disappearance into the complex plane as an external parameter is varied \cite{Kaplan:2009}. In nonrelativistic systems with short-range interactions, Efimov discovered that the spectrum of a three-body system can have a sequence of infinitely many shallow three-body bound states with an accumulation point at the three-body threshold \cite{Efimov:1970}. This phenomena is called the \textit{Efimov effect}. In the case of identical bosons in three spatial dimensions, the spectrum of Efimov states reflects a discrete scaling symmetry with discrete scaling factor $e^{\pi/s_0} \approx22.7$, where $s_0 =1.00624$.

In this article, we use the RG perspective to explicitly show how the loss of conformal invariance happens in Efimov physics in the case of identical bosons as the spatial dimension $d$ is varied. We discuss briefly  the fixed points in the two-body sector in Section~\ref{two-body}. In Section~\ref{three-body}, we discuss the fixed points and the loss of conformal invariance in the three-body sector in more detail. We summarize our results in Section~\ref{conc}.

\section{Two-body Sector}\label{two-body}
In three spatial dimensions,  the low-energy scattering of two nonrelativistic particles with short-range interactions is characterized by the $s$-wave scattering length $a$ only. The \textit{zero-range} limit is defined by taking the range of interaction to zero with the scattering length $a$ held fixed. In the \textit{unitary} limit defined by $a\rightarrow\pm\infty$,  the two-body system is scale invariant. 
From the RG perspective, scale invariance arises at zeros of the beta function, which correspond to  RG fixed points.

We consider the behavior of identical bosons in the zero-range limit in $d$ dimensions. We use the effective field theory (EFT) of Bedaque, Hammer and van Kolck (BHvK) that was used to calculate the behavior of three identical bosons in three spatial dimensions \cite{Bedaque:1999}.  This EFT has a dynamical field $\psi$ and an auxiliary diatom field $\Delta$. The BHvK Lagrangian density is
\begin{equation}
 {\cal L}_{\rm BHvK}=\psi^{\dagger}\left(i\frac{\partial}{\partial t}+\frac{1}{2}\nabla^2\right)\psi+\frac{g_2}{4}\Delta^{\dagger}\Delta-\frac{g_2}{4}\left(
\Delta^{\dagger}\psi^2+\psi^{\dagger 2}\Delta\right)-\frac{g_3}{36}\left(\Delta\psi\right)^{\dagger}\left(\Delta\psi\right),
\label{Lagrangian_three_body}
\end{equation}
where $g_2$ and $g_3$ are the bare two-body and three-body coupling constants. 
The perturbative expansion of the off-shell two-body scattering amplitude in powers of $g_{\rm 2}$ is UV divergent. It can be regularized by imposing a  cutoff $\Lambda$ on the loop momenta. The bare coupling $g_{\rm 2}$ must depend on the cutoff $\Lambda$ to exactly compensate for the $\Lambda$ dependence in the two-body amplitude. The explicit dependence of the two-body coupling $g_2$ on the cutoff $\Lambda$ is given in Eq.~(5) of Ref.~\cite{Mohapatra:2018}. 
We define a dimensionless two-body coupling $\hat{g}_2$ by
\begin{equation}
\hat{g}_2(\Lambda)=\frac{1}{(d-2)(4\pi)^{d/2}
	\Gamma\big(\frac{d}{2}\big)} \Lambda^{d-2} g_2(\Lambda).
\label{dimensionless_g2}
\end{equation}
In terms of the dimensionless coupling $\hat{g_2}$, the RG equation is 
\begin{equation}
\Lambda\frac{d~}{d\Lambda} \hat{g}_2=
(d-2)\hspace{0.1 cm}\hat{g}_2\left(\hat{g}_2 + 1\right).
\label{differential_RG}
\end{equation}
The $\beta$ function  defined by the right hand side of the above equation has zeros at
$\hat{g}_2=-1$ and $\hat{g}_2=0$, which are an IR and a UV fixed point of the RG flow. 
The IR fixed point corresponds to the
trivial non-interacting limit,  
whereas the UV fixed point corresponds to the non-trivial unitary limit. 
There are  scale-invariant interactions in the two-body sector at the UV fixed point. 
\vspace{-0.4 cm}
\section{Three-body Sector}\label{three-body}
\vspace{-0.15 cm}
Discrete scale invariance in Efimov physics can be associated with an RG flow governed by a limit cycle \cite{Bedaque:1999}. The RG equation for the three-body coupling $g_3$ was first derived by Braaten and Hammer in $d=3$ in Ref.~\cite{Eric_Hammer}.  The RG equation for $g_3$ in $d$ spatial dimensions was derived by us in Ref.~\cite{Mohapatra:2018}. The detailed derivation of the results in this section are given in Ref.~\cite{Mohapatra:2018}.

\begin{figure}[t]
\floatbox[{\capbeside\thisfloatsetup{capbesideposition={right,center},capbesidewidth=4.0cm}}]{figure}[\FBwidth]
{\caption{The relevant branch of solutions of Eq.~\eqref{scaling_solution} for $s^2$ as a function of the dimension $d$. 
The value of  $s^2$ decreases from $1$ at $d=2$ and $d=4$ to a minimum value of $-1.016$ at $d=3.04$. }
\label{svsd}}
{\includegraphics*[width=7.5 cm,clip=true]{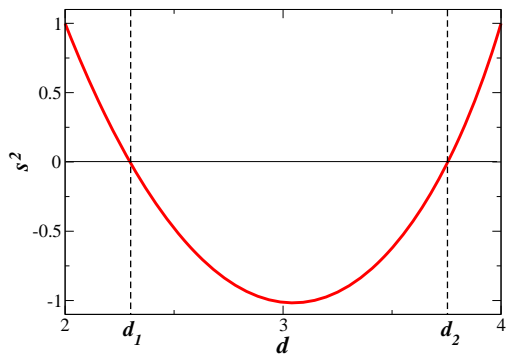} }
\end{figure}

The dependence of the three-body coupling $g_3$ on the UV cutoff $\Lambda$ can be determined from the off-shell atom-diatom amplitude. In the center-of-mass frame, the atom-diatom amplitude is a function of the incoming and outgoing relative momenta ${\bm p}$ and ${\bm k}$ and the total energy $E$. It satisfies the Skorniakov-Ter-Martirosian (STM) integral equation \cite{STM:1957}, which is given explicitly in Eq.~(14) of Ref.~\cite{Mohapatra:2018}. 
The integral equation involves  a dimensionless three-body coupling $\hat{G}(\Lambda)$ defined by
\begin{equation}
\hat{G}(\Lambda)=\frac{\Lambda^2 g_3(\Lambda)}{9g_2(\Lambda)^2} .
\label{dimensionless_H}
\end{equation}
In the limit $\Lambda\rightarrow\infty$, the solutions to the STM equation depend log-periodically on the cutoff $\Lambda$. The dependence of the three-body coupling $g_3$, or equivalently the dimensionless coupling $\hat{G}$, on $\Lambda$ can be determined by demanding that the solutions to the integral equation are well behaved in the limit $\Lambda\rightarrow\infty$. Since $\hat{G}$ is the only coupling in the integral equation, the $\beta$ function for $\hat{G}$ can only depend on $\hat{G}$. The RG equation for the dimensionless coupling $\hat{G}$ is  
\begin{equation}
 \Lambda\frac{d~}{d\Lambda} \hat{G}=\frac{1-s^2}{2}+\left(1+s^2\right)\hat{G}
 +\frac{1-s^2}{2}\hat{G}^2,
 \label{RG_three_fixedpoint}
 \end{equation}
where $s^2$ is a function of dimension $d$ that satisfies a transcendental equation: 
\begin{equation}
2\sin\left(\frac{d}{2}\pi\right){_2}F_1\!
\left({\frac{d-1+s}{2}, \frac{d-1-s}{2} \atop  \frac{d}{2}}\,{\Bigg |}\frac{1}{4}\right)+\cos\left(\frac{ s}{2}\pi\right)=0.
\label{scaling_solution}
\end{equation}
The transcendental equation has infinitely many branches of solutions for $s^2$ as a function of $d$, but the physically relevant branch is shown in Fig.~\ref{svsd}. At $d=3$, the value of $s^2$ is $-s_0^2$, where $s_0=1.00624$ determines  the discrete scale factor $e^{\pi/s_0}$ of Efimov physics. 
There are two critical dimensions for which  Eq.~\eqref{scaling_solution} is satisfied for $s^2=0$:
 \begin{equation}
d_{\rm 1}=2.30,\quad d_{\rm 2}=3.76.
\label{solution_critical_dimension}
\end{equation}
These lower and upper critical dimensions for the Efimov effect were first derived in Ref.~\cite{Nielsen:2001}.

In the regions $d<d_1$ and $d>d_2$, the value of $s^2$ is positive. The $\beta$ function in Eq.~\eqref{RG_three_fixedpoint} for the dimensionless coupling $\hat{G}$ has zeros at $\hat{G}_{\rm +}$ and $\hat{G}_{\rm -}$, which correspond to a UV and an IR fixed point of the RG flow:
 \begin{equation}
\hat{G}_{\rm \pm}=-\frac{1\pm s}{1\mp s}.
 \label{three_body_fixedpoints}
 \end{equation}
As $d$ approaches $d_1$ from below and $d_2$ from above, the value of $s$ approaches $0$, so the two fixed points approach $-1$.

In the region $d_1<d<d_2$, the value of $s^2$ is negative. The $\beta$ function for $\hat{G}$ has complex zeros, which implies that there are no fixed points. The RG flow for $\hat{G}$ is instead governed by a limit cycle:
\begin{equation}
\hat{G}(\Lambda)=-\frac{\cos\left[s_0\log\left(\Lambda/\Lambda_*\right)
	+\arctan s_0\right]}{\cos\left[s_0\log\left(\Lambda/\Lambda_*\right)
	-\arctan s_0\right]},
\label{RGvariable_limitcycle}
\end{equation}
where $\Lambda_*$ is a constant momentum scale and $s_0=\sqrt{-s^2}$. The expression for $\hat{G}$ in Eq.~\eqref{RGvariable_limitcycle} is a log-periodic function of cutoff $\Lambda$ with period $e^{\pi/s_0}$. At unitarity, the two-body sector is scale invariant $(\hat{g}_2=-1)$, but the three-body sector has discrete scale invariance with  discrete scaling factor $e^{\pi/s_0}$.
\vspace{-0.4 cm}
\section{Summary}\label{conc}
\vspace{-0.15 cm}
 In this work, we explicitly showed how conformality is lost in Efimov physics in the case of identical bosons as the spatial dimension $d$ is varied. There are critical dimensions $d_1=2.30$ and $d_2=3.76$ at which conformality is lost \cite{Nielsen:2001}.  For $d<d_1$ and for $d>d_2$, the RG flow has IR and UV fixed points that merge together at $d=d_1$ and at $d=d_2$, respectively. In the region $d_1<d<d_2$, there are no fixed points and the RG flow is governed by a limit cycle with discrete scaling factor $e^{\pi/s_0}$. 
\vspace{-0.5 cm}
\section*{Acknowledgments}
	This research was supported by the National Science Foundation under grant PHY-1607190 and U.S. Department of Energy, Office of Science, Office of Nuclear Physics, under Award Number  DE- FG02-05ER41368.

%
%

\end{document}